\documentclass[prb,twocolumn,superscriptaddress,showpacs,amsmath,amssymb]{revtex4}

\usepackage{graphicx}
\usepackage{latexsym}
\usepackage{amsmath}
\usepackage{amssymb}
\usepackage{amsfonts}
\usepackage{color}
\usepackage{bm}% bold math
\usepackage{verbatim}

\begin{document}
%\begin{multicols}{1}
%%% article title
\title{Majorana states and longitudinal NMR absorption in $^3$He-B film}

\author{M.\,A.\,Silaev}
 \affiliation{Institute for Physics of Microstructures RAS, 603950
Nizhny Novgorod, Russia.}

\date{\today}

%%% abstract
\begin{abstract}
The topological superfluid $^3$He-B supports massless Dirac
spectrum of surface bound states which can be described in terms
of the self-conjugated Majorana field operators.  We discuss here
the possible signature of surface bound states in nuclear magnetic
resonance absorption spectrum in a $^3$He-B film. It is shown that
transitions between different branches of the surface states
spectrum lead to the non-zero absorption signal in longitudinal
NMR scheme when the frequency is larger than the Larmour one.
\end{abstract}

%%% PACS numbers

 \maketitle

\section{Introduction} Recently much attention has been devoted to the
investigation of bound fermion states on surfaces and interfaces
of topological superfluid $^3$He-B. The presence of surface states
in $^3$He-B can be observed through anomalous transverse sound
attenuation
\cite{NagaiImpedance,ImpedanceExp1,ImpedanceExp2,ImpedanceExp3},
surface specific heat measurements \cite{SpecificHeat}. Gapless
fermion states are supported by the non-zero value of the
topological invariant in $^3$He-B \cite{TopInvHe} and have two -
dimensional relativistic massless Dirac
spectrum\cite{NagaiAnisotropy,SalomaaVolovik1988,
DiracCone,SukBum,Machida}. Such massless fermions can be described
in terms of the Majorana self-conjugated field operators which
have been intensively studied recently in a number of condensed
matter systems \cite{Wilcek}.

It has been suggested that NMR technique can be employed to study
the spectrum of surface states\cite{VolovikIsing}. The NMR
measurements were implemented recently on superfluid $^3$He films
demonstrating various frequency shifts associated with the order
parameter dynamics both at high-temperature $A$ phase as well as
with two non-equivalent low temperature $B$ phase states
characterized by the different values of the Ising
variable\cite{Levitin1,Levitin2}.

In this paper we focus on the contribution of the surface bound
states to the ac magnetic susceptibility of He$^3$ B film. We
demonstrate that the transitions between different branches of
surface states spectrum result in dissipation manifested in
non-zero imaginary part of the $\chi_{zz}$ component of magnetic
susceptibility, where $z$ is the axis normal to the film surface.
This effect should provide the contribution to the longitudinal
NMR absorption signal absent both in the normal state and bulk
$^3$He B phase. Under the action of the $z$ component of magnetic
field which destroys the self-conjugating Majorana states the
magnetic absorption is suppressed at the frequencies smaller than
the Larmour one.

\section{Spectrum of surface states in $^3$He-B film} At first we
introduce the basic formalism
 for treating the spectrum of fermionic
quasiparticles.  We consider $^3$He-B film confined in a slab at
$z>0$ and homogeneous in $x$ and $y$ directions.  The $^3$He-B
surface mode is derived from the quasiclassical BdG Hamiltonian
 \begin{equation}\label{Eq:Hamiltonian}
 \hat H=-i\hbar V_z\hat\tau_3\partial_z+\hat\tau_1\hat\Delta-\frac{\gamma}{2} {\bf
H}\cdot {\mbox{\boldmath$\hat \sigma$}},
 \end{equation}
 where $\hat\Delta=A_{ij}q_i\hat\sigma_j$, and ${\bf q}={\bf k}/k_F$
where $\hat\tau_i$ are Pauli matrices of Bogolyubov-Nambu spin,
$\hat\sigma_\alpha$ are Pauli matrices of
 $^3$He nuclear spin, $\gamma$ is the gyromagnetic ratio of the $^3$He
 atom.

 The order parameter in $^3$He-B is $3\times3$ matrix
$A_{\alpha i}$ where the Greek and Latin indices correspond to the
spin and orbital variables. The geometrical confinement induced by
the walls destroys the isotropy of the B phase order
parameter\cite{VollhardtWolfle}. As a result For the distorted
3He-B it has the form
 $$
 A_{ij}=\begin{pmatrix}
   \Delta_\parallel & 0 & 0 \\
   0 & \Delta_\parallel & 0 \\
   0 & 0 & \Delta_\perp \
 \end{pmatrix}
 $$
where $\Delta_\perp$ is the gap  for quasiparticles propagating in
direction along the normal to the wall and $\Delta_\parallel$  is
the gap for quasiparticles propagating in directions parallel to
the wall.

 Upon the specular reflection from the surface the quasiparticle momentum projection $q_z$
 and therefore the part of the order parameter changes the sign which leads to the formation of
 surface bound states. To find the spectrum of this states we employ the usual procedure,
 considering the two parts of hamiltonian
 $\hat H=\hat H_0+\hat H_1$ so that
\begin{eqnarray}\label{Eq:H0}
 \hat H_0=-i\hbar V_z\hat\tau_3\partial_z+\hat\tau_1 \hat F(z) \\
 \label{Eq:H1}
\hat H_1=\Delta_\parallel\hat\tau_1\left(\hat\sigma_x
 q_x+\hat\sigma_y q_y\right)-\frac{\gamma}{2} {\bf H}\cdot
 {\mbox{\boldmath$\hat\sigma$}},
\end{eqnarray}
 where $\hat F(z)=\Delta_\perp (z)\hat\sigma_z q_z$ and ${\bf q}={\bf p}/p_F$
 The hamiltonian (\ref{Eq:H0}) has zero energy eigenvalues
corresponding to the degenerate surface bound states.
 To get into account correction from the perturbation terms $\hat H_1$
 we find the eigenfunctions as a superposition
 \begin{equation}\label{solTB2}
 \psi=\sum_{j=1}^2 X_j \varphi_{j}(z) ,
 \end{equation}
 where $X_{k}$ are the arbitrary coefficients and the generic terms $\varphi_{j}(z)$ are
 the eigenfunctions of hamiltonian $\hat H_0$
corresponding to the zero energy $\varepsilon=0$
\begin{equation}\label{Eq:ZeroOrderFunction}
\varphi_{j}(z)=A^{-1/2} \alpha_j\beta_j \exp\left[-K(z)\right],
\end{equation}
 where $A=\langle\varphi_1|\varphi_1\rangle=\langle\varphi_2|\varphi_2\rangle$ is the normalizing coefficient
 and
 \begin{equation}%\label{Kv}
 \nonumber
 K(z)=\frac{1}{\hbar V_F}\int_{0}^z \Delta_\perp(s) ds.
 \end{equation}
The Pauli and Nambu spinors $\beta_j$ and $\alpha_j$ satisfy the
relations $\hat\sigma_z\beta_{1,2}=\mp \beta_{1,2}$ and
 $\hat\tau_2\alpha_{1,2}=\pm\alpha_{1,2}$.

 Following the standard method we substitute the
 solution in the form (\ref{solTB2}) into the equation $(\hat
 H_0+\hat H_1)\psi=\varepsilon\psi$ multiply by $\psi^*_{j}(z)$ from the left and
 integrate over $z$. Then for the spinor $X= (X_1, X_2)^T$ we
 obtain the two-dimensional Dirac equation
\begin{equation}\label{Eq:Dirac}
\left[C{\bf {\mbox{\boldmath$\hat\sigma$}} p} +\hat\sigma_z
M\right] X= \varepsilon X,
\end{equation}
 with the 'light velocity' given by
  $$
 C=\frac{1}{Ap_F}\int_0^{\infty}dz \Delta_\parallel (z)
 \exp\left[-2K(z)\right]\sim
\Delta_\parallel/p_F
 $$
and 'mass' determined by the Larmour frequency
$M=\hbar\omega_H/2$.

For the massless particles we can choose the eigenfunctions of
Eq.(\ref{Eq:Dirac}) satisfying the relation
\begin{equation}\label{Eq:SelfConjugate}
X_{\varepsilon}=i\hat\sigma_y X_{-\varepsilon}.
\end{equation}
As a result the quasiparticle field operators are self-conjugated
analogously to the Majorana fermions in relativistic quantum field
theories. In the external magnetic field the particles described
by Eq.(\ref{Eq:Dirac}) become massive\cite{VolovikIsing} so that
the property (\ref{Eq:SelfConjugate}) does not hold and therefore
the quasiparticles are no longer self-conjugated Majorana
fermions.

  The Dirac equation (\ref{Eq:Dirac}) determines the equation for the energy
  levels in the following form\cite{VolovikIsing}
  \begin{equation}\label{Eq:Energy}
  \varepsilon_{1,2}=\pm \sqrt{(Cp)^2+(\hbar\omega_H)^2/4}.
  \end{equation}
  The energy spectrum of surface bound states (\ref{Eq:Energy}) is
  sensitive only to the $z$ projection of the magnetic field.
  It can result in a large anisotropy of magnetic susceptibility\cite{NagaiAnisotropy}
  if the magnetic field is much smaller then the effective dipole
  field. However the larger magnetic field will reorient the
  spin axes eliminating the magnetic anisotropy\cite{VolovikIsing}.
  Note that deriving the spectrum (\ref{Eq:Energy}) we have
  neglected the finite thickness of the slab. The size effect due to the overlap of quantum
  states localized at the opposite surfaces of the slab leads to
  the splitting of Dirac cone\cite{Machida} even in zero magnetic
  field. For sufficiently strong magnetic field this modification
  can be neglected.

 \section{Magnetic susceptibility and NMR absorption spectrum}
    Now we consider the contribution of surface bound states
  to the imaginary part of ac magnetic susceptibility component
  $\chi_{zz}$ which determines the power absorption under the
   experimental conditions of the longitudinal
  scheme of magnetic resonance when both the total magnetic field is
  directed along the $z$ axis.

To find the magnetic susceptibility let us use a conventional Kubo
formula:
\begin{equation}\label{Eq:Kubo}
\chi_{ij}= \frac{\gamma^2}{4}T \sum_{\omega_n} \int \frac{d^2 {\bf
p}}{(2\pi\hbar)^2} Tr \{\hat\sigma_i \hat G (p_+)\hat\sigma_j \hat
G(p_-)\},
\end{equation}
 where
 $$
 \hat G (\omega_M,{\bf p}) =\sum_{k=1,2}
  \frac{|\psi_k\rangle \langle\psi_k|}{i\omega_M-\varepsilon_{k}({\bf p})}
 $$
 is a temperature Green function and $p_{\pm}=(\omega_n\pm \epsilon_l/2, {\bf
 p})$. Here $\omega_n=\pi (2n+1) T$ is a fermionic and $\epsilon_l=2\pi l T$ is a
 photonic Matsubara frequencies. The normalized wave functions
 $\psi_{1,2}$ are given by the superpositions (\ref{solTB2}) and
 correspond to the energy branches $\varepsilon_{1,2}$ in
 Eq.(\ref{Eq:Energy}). We use the value of the matrix element
 $|\langle\psi_1|\check{\sigma}_z|\psi_2\rangle|=Cp/\varepsilon_1$
 and the formula for the sum over fermionic frequencies
 $$
 \sum_{\omega_n}\frac{T}{[i\Omega_{n+}-\varepsilon_1][i\Omega_{n-}-\varepsilon_2]}
  =\frac{f_0(\varepsilon_1)-f_0(\varepsilon_2)}
 {i\epsilon_l-\varepsilon_2+\varepsilon_1}.
 $$
 where $\Omega_{n\pm}=\omega_n\pm\epsilon_l/2$ and
 $f_0(\varepsilon)=\tanh(\varepsilon/2T)$. Then changing the photonic Matsubara frequency
  by $\epsilon_l\rightarrow i\hbar\omega$ from Eq.(\ref{Eq:Kubo})we get that
\begin{equation}\label{Eq:ZZ1}
 \chi_{zz}=\frac{\gamma^2C^2}{8\pi\hbar^2} \int p^3d p
 \frac{(\varepsilon_2-\varepsilon_1)\left[f_0(\varepsilon_1)-f_0(\varepsilon_2)\right]}
 {\varepsilon_1^2[(\hbar\omega)^2-(\varepsilon_2-\varepsilon_1)^2]}
  \end{equation}
 where $\varepsilon_{1,2}=\varepsilon_{1,2}({\bf p})$.
We use now the dispersion relation (\ref{Eq:Energy}) and obtain
that when the  frequency is larger than the Larmour frequency
$\omega>\omega_H$
 the susceptibility given by Eq.(\ref{Eq:ZZ1}) has a non-zero
 imaginary part
 \begin{equation}\label{Eq:ImZZ}
  Im \chi_{zz}=\left(\frac{\gamma}{4\hbar C}\right)^2\frac{\hbar\omega}{2}f_0\left(\frac{\hbar\omega}{4T}\right)
  \left(1-\frac{\omega_H^2}{\omega^2}\right).
 \end{equation}
   Note that the non-zero dissipation $Im \chi_{zz}\neq 0$ occurs
    only when the frequency is larger than the threshold
 value $\omega_H$ which is similar to the threshold behavior of
 absorption rate in semiconductors where the absorption edge frequency is determined by the band gap
 energy. However in the vicinity of the threshold the frequency dependence of
 absorption rate Eq.(\ref{Eq:ImZZ}) is completely different from
 that of the electromagnetic wave absorption in the
 physics of semiconductors\cite{Semiconductors}.

 The estimate of imaginary magnetic susceptibility (\ref{Eq:ImZZ})
 of a unit film area yields
 $$
 Im \chi_{zz}\sim \chi_n \xi \frac{\hbar\omega}{\Delta_\parallel}f_0\left(\frac{\hbar\omega}{4T}\right)
 $$
where $\chi_n\sim \gamma^2k_F^3/E_F$ is a normal state
susceptibility and $\xi=\hbar V_F/\Delta_\parallel$ is a coherence
length.

\section{Effect of surface roughness} In general the quasiparticle
energy levels are broadened due to the statistical fluctuation of
the film surface which affect the boundary conditions for the wave
functions. Different models of surface roughness related to the
surface effects in $^3$He were developed including the diffusive
surface layer\cite{DiffusiveLayer,GeneralBC}, randomly rippled
wall (RRW), randomly oriented mirrors (ROM)
models\cite{GeneralBC,ROM} and random scattering matrix model
\cite{RandomScatteringMatrix}. Under the conditions of diffusive
scattering the acoustic impedance data demonstrate the presence of
surface bound states in $^3$He-B film
\cite{ImpedanceExp1,ImpedanceExp2} although there is no evidence
of the relativistic massless Dirac spectrum. On the other hand the
surface conditions can be varied in the experiments by coating the
surface of several layers of $^4$He\cite{He4Layer}. For increased
specularity factor the new features on the temperature dependence
of acoustic impedance were observed indicating the formation of 2D
Dirac energy spectrum\cite{ImpedanceExp1}.

Here we employ the ROM model assuming that the surface consists of
small randomly oriented specularly scattering facets\cite{ROM}.
This model is applicable to describe the fluctuations with the
scale much larger than $k_F^{-1}$ of the $^4$He coated surface of
the film. Within the framework of ROM model the important
characteristic is the angle $\alpha_s$ which constitutes the local
normal vector to the wall ${\bf n_s}$ with the $z$ axis. Let us
use the new coordinate system rotated by the angle $\alpha_s$ with
respect to the axis defined by $\nu={\bf z}\times {\bf n_s}$. Then
we obtain in the new coordinate system the expression for the
order parameter matrix $\tilde{A}_{ij}=A_{ik}R_{kj}$, where $\hat
R$ is the corresponding rotation matrix. Let us assume without
loss of generality that the rotation axis coincide with the $y$
axis. Then in the Eq.(\ref{Eq:H0}) for the hamiltonian $\hat H_0$
we obtain
 $\hat F=(\Delta_\perp \cos\alpha_s\hat\sigma_z-\Delta_\parallel \sin\alpha_s\hat\sigma_x)q_z$
and the perturbation term is given by
 \begin{equation}\label{Eq:H11}
 \hat
 H_1=\hat\tau_1q_x\left(\Delta_\parallel\hat\sigma_x\cos\alpha_s
 +\Delta_\perp\hat\sigma_z\sin\alpha_s \right)-\frac{\gamma}{2} {\bf H}\cdot
 {\mbox{\boldmath$\sigma$}}.
 \end{equation}

To proceed further with analytical calculations we assume that the
order parameter does not depend on the space coordinates so that
$\Delta_\perp, \Delta_\parallel= const$. Then we obtain easily the
zero energy eigenvectors of the hamiltonian $\hat H_0$ in the form
of Eq.(\ref{Eq:ZeroOrderFunction}) with
 \begin{align}\label{}
 &\beta_1=\left(\frac{\Delta_\parallel\sin\alpha_s}{\Delta_\perp\cos\alpha_s+\bar{\Delta}},
 1\right)^T
 \\
 &\beta_2=\left(1,-\frac{\Delta_\parallel\sin\alpha_s}{\Delta_\perp\cos\alpha_s+\bar{\Delta}}\right)^T
\end{align}
 where
 $\bar{\Delta}=\sqrt{(\Delta_\parallel\sin\alpha_s)^2+(\Delta_\perp\cos\alpha_s)^2}$.
 Correspondingly in the Eq.(\ref{Eq:ZeroOrderFunction}) for the
 zero order wave functions we obtain
 \begin{equation}%\label{Kv}
 \nonumber
 K(z)=\frac{1}{\hbar V_F}\int_{0}^z \bar{\Delta} ds.
 \end{equation}
 The quasiparticle spectrum obtained along the perturbation theory
 scheme described above yields the spectrum in the form (\ref{Eq:Energy})
 but with modified parameters.
 We will study the modification of the absorption threshold which
 is determined by the 'mass' term and does not depend on the
 'light velocity'. We therefore will neglect
 the modification of 'light velocity' and focus on the 'mass' term
 which is given by
 \begin{equation}\label{Eq:mass}
    \tilde{\omega}_H=\omega_H\frac{(\Delta_\perp\cos\alpha_s+\bar{\Delta})^2-(\Delta_\parallel\sin\alpha_s)^2}
 {(\Delta_\perp\cos\alpha_s+\bar{\Delta})^2+(\Delta_\parallel\sin\alpha_s)^2}.
\end{equation}
To proceed further and calculate statistical average
 over the surface roughness we assume that $|\alpha_s|\ll 1$ and
 obtain to the leading order
\begin{equation}\label{Eq:OmegaHfluct}
\tilde{\omega}_H=\omega_H\left(1-\frac{\alpha_s^2\Delta_\parallel^2}{2\Delta_\perp^2}\right).
\end{equation}
The above Eq. yields the fluctuating correction to the absorption
edge in Eq.(\ref{Eq:ImZZ}). It leads to the smoothing out the
sharp absorption edge at the Larmour frequency. To estimate this
effect we assume the Gaussian distribution of angle $\alpha_s$
with the zero average value $\langle\alpha_s\rangle=0$ and the
dispersion $\langle\alpha_s^2\rangle=\sigma_\alpha^2$. After that
the average value of the susceptibility in the vicinity of Larmour
frequency is given by $$
 Im \chi_{zz}=\left(\frac{\gamma}{4\hbar C}\right)^2\left(\frac{\Delta_\parallel}{\Delta_\perp}\right)^2
 \frac{\hbar\omega}{2}f_0\left(\frac{\omega}{4T}\right)S(\alpha_0,\sigma_\alpha)
$$
where
 $\alpha_0=2\left(\Delta_\perp/\Delta_\parallel\right)^2
 \left[1-\left(\omega/\omega_H\right)^2\right]$  and
\begin{equation}\label{Eq:S}
S(\alpha_0,\sigma_\alpha)=\int_{\alpha^*}^\infty
(\alpha^2-\alpha_0)\frac{\exp\left[-(\alpha/\sigma_\alpha)^2\right]}{\sigma_\alpha\sqrt{\pi}}
 d\alpha
\end{equation}
where $\alpha^*=\sqrt{\alpha_0}$ if $\alpha_0>0$ and $\alpha^*=0$
if $\alpha_0<0$.

\begin{figure}[hbt]
\centerline{\includegraphics[width=1.0\linewidth]{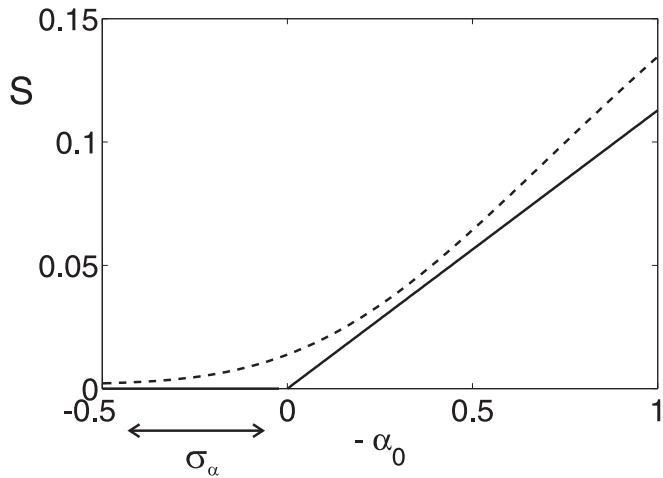}}
\caption{\label{plots}
 Plot of the function $S(\alpha_0,\sigma_\alpha)$ for $\sigma_\alpha=0.01$ (solid line)
 and $\sigma_\alpha=0.7$ (dashed line). }
\end{figure}

 The plots of the
function (\ref{Eq:S}) for the different values of $\sigma_\alpha$
are shown in Fig.(\ref{plots}) demonstrating smoothing of
absorption edge with increasing dispersion of surface ripples.
Although in general  for $\sigma_\alpha>0$ the absorption signal
is non-zero at the whole frequency domain however it is
exponentially decaying for $\omega\ll\omega_H$. The size of the
crossover domain in Fig.(\ref{plots}) is determined by the
dispersion $\delta\omega=\omega_H\sigma_\alpha$.
 Therefore in general we can conclude that the
absorption edge should well observed provided these fluctuations
are small so that $\sigma_\alpha\ll 1$.

%%%%%%%%%%%%%%%%%%%% Figure %%%%%%%%%%%%%%%%
%%%%%%%%%%%%%%%%%%%%%%%%%%%%%%%%%%%%%%%%%%%%

\section{Conclusion} To conclude we have calculated the contribution of
fermionic surface bound states to the ac magnetic susceptibility
of $^3$He-B film. We have shown that in the longitudinal NMR
scheme the non-zero absorption signal appears provided the
frequency is larger than the threshold one determined by the
Larmour frequency $\omega>\omega_H$. Such absorption is absent in
the normal state of $^3$He and can not occur due to the dynamics
of the order parameter spin either. In zero magnetic field there
is no frequency threshold for the dissipation which can be
considered as the fingerprint of the gapless Majorana surface
bound states.
 The surface fluctuations are shown to smooth the threshold behavior
out providing the small absorption in the frequency domain
$\omega<\omega_H$.

\section{Acknowledgements}
 This work was supported, in part, by ``Dynasty'' Foundation, Russian Foundation for Basic Research,
 Presidential RSS Council (Grant No. MK-4211.2011.2),
 by Programs of RAS ``Quantum Physics of Condensed Matter'' and ``Strongly
 correlated electrons in semiconductors, metals, superconductors and magnetic
 materials''. Discussions with G.E. Volovik and A.S.
Mel'nikov are greatly acknowledged.


\begin{thebibliography}{99}

   \bibitem{NagaiImpedance}
 K. Nagai, Y.Nagato, M. Yamamoto, S. Higashitani,
  J. Phys. Soc. Japan,  {\bf 77}, 111003
 (2008);

 \bibitem{ImpedanceExp1}
 S. Murakawa, Y. Tamura, Y. Wada, M. Wasai, M. Saitoh, Y. Aoki, R. Nomura, Y. Okuda, Y. Nagato,
M. Yamamoto, S. Higashitani, and K. Nagai, Phys. Rev. Lett. {\bf
103}, 155301 (2009)

\bibitem{ImpedanceExp2}
J. P. Davis, J. Pollanen, H. Choi, J. A. Sauls, W. P. Halperin, A.
B. Vorontsov, Phys. Rev. Lett. {\bf 101}, 085301 (2008)

\bibitem{ImpedanceExp3}
Y. Aoki, Y. Wada, M. Saitoh, R. Nomura, Y. Okuda,  Y. Nagato, M.
Yamamoto,  S. Higashitani, K. Nagai, Phys. Rev. Lett. {\bf 95},
075301 (2005).

\bibitem{SpecificHeat}
H. Choi, J. P. Davis, J. Pollanen, and W. P. Halperin, Phys. Rev.
Lett. {\bf 96}, 125301 (2006)

  \bibitem{NagaiAnisotropy}
  Y.Nagato, S. Higashitani, and K. Nagai, J. Phys. Soc. Japan, {\bf 78},123603 (2009);

   \bibitem{TopInvHe}  G.E. Volovik,
   JETP Lett. {\bf 90}, 587 (2009);

   \bibitem{SalomaaVolovik1988}
 M.M. Salomaa and  G.E. Volovik,
 Phys. Rev. B {\bf 37}, 9298 (1988).

    \bibitem{DiracCone}
 G.E. Volovik, JETP Lett. {\bf 90}, 398 (2009);

 \bibitem{SukBum}
Suk Bum Chung, Shou-Cheng Zhang, Phys. Rev. Lett. {\bf 103},
235301 (2009).

 \bibitem{Machida}
 Y. Tsutsumi, M. Ichioka, and K. Machida,
 Phys. Rev. B {\bf 83}, 094510 (2011).

 \bibitem{Wilcek}
 F. Wilczek,
 Nat. Phys. {\bf 5}, 614 (2009).

\bibitem{VolovikIsing}
G.E. Volovik, JETP Letters, {\bf 91}, 215 (2010);

 \bibitem{Levitin1}
 R.G. Bennett, L.V. Levitin, A. Casey, B. Cowan, J.Parpia,
 J. Saunders, J. Low. Temp. Phys.,  {\bf 158}, 163
 (2010);

 \bibitem{Levitin2}
 L.V. Levitin, R.G. Bennett, A. Casey, B. Cowan, J.Parpia,
 J. Saunders, J. Low. Temp. Phys.,  {\bf 158}, 159
 (2010);

\bibitem{VollhardtWolfle}
D. Vollhardt and  O. W\"olfle, {\it The superfluid phases of
helium 3},  Taylor and Francis, London  (1990).

\bibitem{Semiconductors}
R.A. Smith, {\it Semiconductors}, Cambridge University Press
(1961).

\bibitem{DiffusiveLayer}
N.B.Kopnin, P.I.Soininen, and M.M. Salomaa, J.Low Temp. Phys {\bf
85}, 267 (1991).

\bibitem{GeneralBC}
L.J. Buchholtz and D.Rainer, Z.Phys B {\bf35} 151 (1979)

\bibitem{ROM}
E. V. Thuneberg, M. Fogelstrom and J. Kurkij/irvi, Physica
 {\bf B178} 176 (1992).

\bibitem{RandomScatteringMatrix}
Y.Nagato, S. Higashitani, K.Yamada, and K.Nagai, J.Low Temp. Phys
{\bf 103}, 1 (1996);

\bibitem{He4Layer}
 S. M. Tholen and J. M. Parpia, Phys. Rev. Lett. {\bf 68}, 2810
 (1992); D.Kim et.al, Phys. Rev. Lett. {\bf 71}, 1581, (1993);
 M.R.Freeman and R.C.Richardson, Phys. Rev. B {\bf 41}, 11011, (1990).

%
%\bibitem{Falkovskii}
% L.A. Falkovsky, A.A. Varlamov, Eur.Phys. J. B {\bf 56}, 281 (2007)
%
%\bibitem{NMRUllah}
% S. Ullah, Phys. Rev. B {\bf 37}, 5010
% (1987);
%
%\bibitem{NMRBrinkman}
% W.F. Ârinêman, Phys. Lett. A, {\bf 49}, 411, (1974)


\end{thebibliography}
\end{document}